# Mechanical analysis of pimple growth and pain level characterization

Xiangbiao Liao[1], Xiaobin Deng[2], LiangLiang Zhu[2], Feng Hao[1], Hang Xiao[1], Xiaoyang Shi[1] and Xi Chen[1,2,*]

[1]Department of Earth and Environmental Engineering, Columbia University, New York, NY 10027, USA

[2]International Center for Applied Mechanics, State Key Laboratory for Strength and Vibration of Mechanical Structures, School of Aerospace, Xi'an Jiaotong University, Xi'an 710049, P.R. China

[*]Corresponding author: xichen@columbia.edu

## Abstract

Pimple is one of the most common skin diseases for humans. The mechanical modeling of pimple growth is very limited. A finite element model is developed to quantify the deformation field with the expansion of follicle, and then the mechanical stimulus is related to the sensation of pain during the development of pimple. Through these models, parametric studies show the dependence of mechanical stimulus and pain level on the pimple-surrounded structures, follicle depth and mechanical properties of the epidermis. The findings in this paper may provide useful insights on prevention or pain relief of pimples, as well as those related to cosmetics and other tissue growth.

## 1. Introduction

Skin serves as the first line of defense against external stimuli and pathogens invasion, and also plays a key role in chemicals regulation [1]. The integrity of skin is not only essential for physiological health but also crucial for personal appearance. Unfortunately, a few skin diseases, for example, acne vulgaris as the worldwide common disease, usually result in physical and even psychological suffering [2]. Pimple, one type of acne vulgaris and the red bumps along with a white core, usually flourishes among the one during puberty stage, and the resulting pain and appearance with scarring can cause the reduction of self-esteem and even depression [3,4].

It was shown that the drama of pimple is played out in the sebaceous follicle [2]. Excessive sebum produced by sebaceous glands is clogged up in the follicle due to the blockage of hair pores by dead skin cells, leading to the formation of bump and then the inflammation caused by bacteria harboring [2,5,6]. It's observed that the forehead is usually the first place to be involved for acne growth [2]. While there are a few treatments that can reduce the lesion of acnes, including antibiotics medications and hormonal treatments [4,7], there has been no ideal treatment for pimples so far [8]. Most of previous studies of pimple are based on anatomy, clinical observations, or questionnaire-based observations [2-10], however mathematical descriptions of the pimple growth are limited. As the excessive secreted sebum is responsible for the formation of convex bump during the development of acne [11], understanding the interaction between sebum and surrounding components of skin is necessary.

Mechanical principles have been employed to provide useful insights to a few biological phenomena. For example, the morphologies of a broad set of natural and biological systems, such as Korean melons, pumpkins, ridged gourds [12] and even wrinkled fingertips [13], are highly relevant to the elastic buckling of a thin film on a compliant substrate system. In addition, mechanical models based on anatomical structures of eyelid are fit for explaining various morphologies of eyelids, which may further guide surgeries related to the eyelid (e.g., double-eyelid surgery) [14]. It is interesting to see that mechanical principles may play a critical role in cosmetology. Therefore, exploring growth mechanics of pimple potentially inspires medical practitioners to gain insight into the preventative treatments of acnes.

In addition, pimple-induced pains are remarkably common for humans in everyday life causing intensive facial lesions, and usually are induced by the local mechanical stimulus from expansion of sebum under the skin and bacteria inflammation [2]. Moreover, it is commonsensible that pimples grown on different body sites evoke variable sensation of pain. For example, forehead pimples usually induce higher level pain than those on the cheek. However, mathematical modeling of the pimple-induced pain is still in demand. By converting the mechanical stimulus into electrical signals via nerve impulses, mechanoreceptors in skin play a key part in pain sensation [15]. Lu, et al. have developed a holistic method to model skin thermal pain [16-19], but there is a lack of direct modeling to relate mechanical deformation at the location of pimple bump to the sensation of pain.

In this study, we investigate the growth of pimple from the perspective of mechanics and build a quantitative bridge between the deformation of pimple bump and sensation of pain. Finite element models based on different sophistication levels of skin anatomical structures are developed to simulate the deformation surrounding pimples and the holistic model is used to characterize the level of pain. We explore the deformation and evoked pain of pimples grown on different body sites by varying geometrical parameters and boundaries. By modification of the mechanical properties of epidermis, the pimple-related pain can be altered, implying a possible cosmetic method to release the level of pain.

## 2 Methods

The schematic of pimple in Figure 1(a) shows that the excessive secreted sebum deforms the follicle in the dermis due to the pore blockage [2] and sequentially cause the skin to elevate in the form of a bump. Since the size of hair is much smaller than that of the follicle and hair pore is blocked [20], the sebum is assumed to be sealed in the follicle void. Accordingly, the accumulation of liquid sebum and development of pimple herein are modeled via applying a uniform pressure to the surface of follicle hole embedded in skin while for simplicity, the plane-strain model is carried out as shown in Figure 1(b). The initial diameter of follicle is assumed to be 0.4 mm. For simplicity, we treat the skin as a laminated structure composed of the layers of epidermis and dermis, which are modeled with an isotropic and linear stress-strain relationship, and the relevant parameters of skin refer to Table 1. A non-linear finite element method, based on the commercial software ABAQUS, is adopted to explore the quasi-static deformation of developing pimples. The boundaries $AB$ and $CD$ are set to be fixed, and their effects on the local deformation of pimple can be neglected due to the significant lateral size (L = 10 mm). The fixed boundary $BD$ is adopted to simulate the case of pimple grown on the forehead with less subcutaneous fat, while a soft substrate, with infinite thickness, Young's modulus 0.01 MPa, Poisson's ratio 0.48, is added below the dermis for the case of pimple grown on soft parts, for example, the cheek.

The holistic method is used to explore the perception of nociceptive pain evoked by the deformation of skin around pimples [17,19]. First, the mechanoreceptors are triggered by mechanical stimulus, resulting in the current in ion channels of nociceptors. A linear relationship between the generated current and the mechanical stimulus $\sigma_p$, is assumed as

$$I_{mech} = C_m(\sigma_p - \sigma_t)/\sigma_t \tag{1}$$

where $\sigma_t = 0.2\text{MPa}$ [21] denotes the mechanical pain threshold and $\sigma_p$ is assumed to be the stress at the location of mechanical nociceptors. Sequentially, the triggered current evokes the generation of membrane potential $V_{mem}$ in the mechanoreceptors, which could be modeled by Hodgkin-Huxley model of nerve excitation:

$$C_{mem}\frac{dV_{mem}}{dt} = I_{mech} + g_{Na}m^3h(E_{Na} - V_{mem}) + g_K n^4(E_K - V_{mem}) + g_L(E_L - V_{mem}) \tag{2}$$

in which $E_{Na} = 55$ mV, $E_K = -72$ mV and $E_L$ are the reversal potentials for sodium, potassium and leakage current components, respectively; $g_{Na} = 120$ mS/cm², $g_K = 36$ mS/cm², $g_L = 0.3$ mS/cm² correspond to the maximum ionic conductance through the three current components; $C_{mem} = 1.0\ \mu F/cm^2$ denotes the membrane capacity per unit area; $t$ is time; the gating variables $m,\ n,\ h$ are dependent on the membrane voltage [19,22,23].

Once mechanoreceptors are triggered, the frequency of membrane potential is transmitted from the skin to spinal cord and brain, while the amplitude of signals is not carried. We employ the gate control theory [24] to model the pain cause by pimple growth:

$$\tau_i \dot{V}_i = -(V_i - V_{i0}) + g_{li}(x_l) + g_{mi}(x_m) \tag{3}$$

$$\tau_e \dot{V}_e = -(V_e - V_{e0}) + g_{se}(x_s, V_e) \tag{4}$$

$$\tau_t \dot{V}_t = -(V_t - V_{t0}) + g_{st}(x_s) + g_{lt}(x_l) + g_{et}(x_e) - g_{it}(x_i) - g_{mt}(x_m) \tag{5}$$

$$\tau_m \dot{V}_m = -(V_m - V_{m0}) + g_{tm}(x_t) \tag{6}$$

where $V_j$ (j = i, e, t, m) denotes membrane voltages of inhibitory SG cell, excitory SG cell, T-cell and midbrain; $V_{j0}$ corresponds to the initial potential; $\tau_j$ is time constant; $x_j$ is firing frequency; $g_{jk}$ describes the potential effect of input ($j$) to a cell ($k$); $x_l$ and $x_s$ respectively represent the frequencies of neural signals transmitted along large and small fibers. The relevant parameters refer to the previous works [16,17]. Since the stretching stimuli in this study are mainly received by one kind of mechanoreceptors, free nerve endings connected with small fibers, $x_l$ is set to zero. The fact of the perception of pain in direct bearing on the potential output from T-cell allows the adoption of the value of $V_t$ at steady state to characterize the pain level during pimple growth [16].

| | Epidermis | 102 |
|---|---|---|
| Young's modulus (MPa) | Dermis | 10.2 |
| Poisson's ratio | Epidermis/Dermis | 0.48 |
| | Epidermis | 0.1 |
| Thickness (mm) | Dermis | 1.5 |

**Table 1.** Geometrical and Mechanical properties of skin

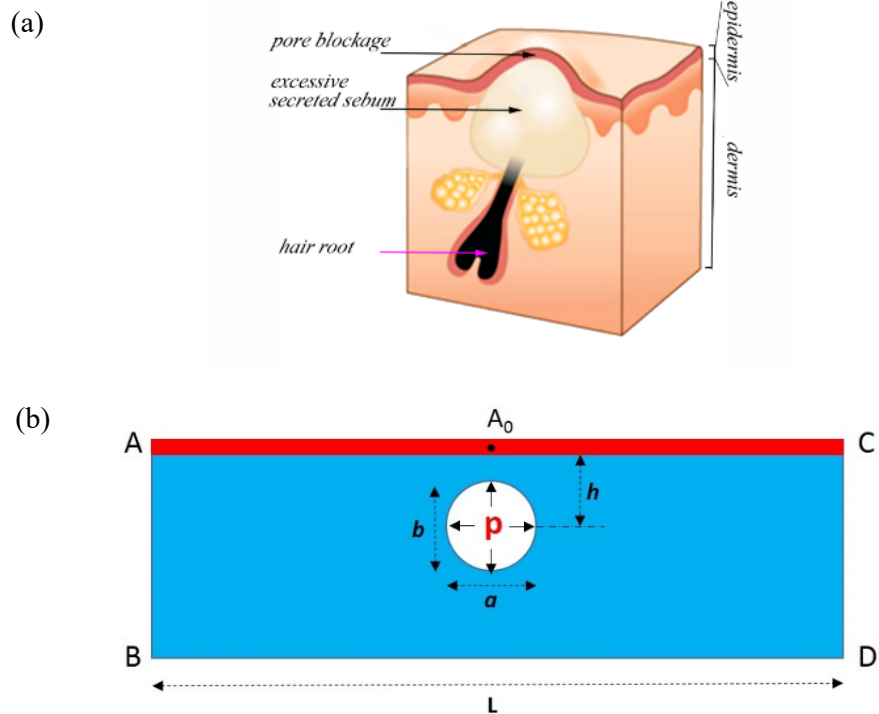

**Figure 1.** Schematics of (a) skin pimple and (b) plane-strain model: $a$ and $b$ denotes the size of the follicle; $h$ and $L$ respectively represents the depth of follicle and lateral size of the model; a uniform pressure, $p$, is applied to the surface of follicle.

**3 Results and Discussion**

**3.1 Deformation and pain level during pimple growth**

Using a case of the follicle originally located in the middle of dermis layer, we explore the deformation field of a growing pimple as secreted sebum (and some other substances) accumulates. The elevated height of pimple bump is characterized by the out-plane displacement of point $A_0$. The inset of Figure 2(a) illustrates the deformation configuration of the pimple with hard substrate. A protruded bump forms on the surface of the model and the size of the enlarged hole is approximately $D = 0.77$ mm, characterized by the average of $a$ and $b$. The simulated configuration seemingly matches with that of real pimple before the inflammatory stage shown in Figure 1(a). As shown in Figure 2(a) and 2(b), it is clearly noticed that the elevated height and maximal principle stress of point $A_0$ increase as the follicle expands, resulting from the increasing volume of blocked sebum under the skin before pimple rupturing. Therefore, the stretching stimulus to skin increases with pimple developing up. Additionally, it's seen from Figure 2(a) and 2(b) that hard substrate induces larger deformation and stress compared to those of the soft substrate. This provides a possible explanation for the common observation of larger pimples on the forehead (hard substrate) than those on the cheek (soft substrate) and that the forehead is usually the first place to be observed of pimple [2].

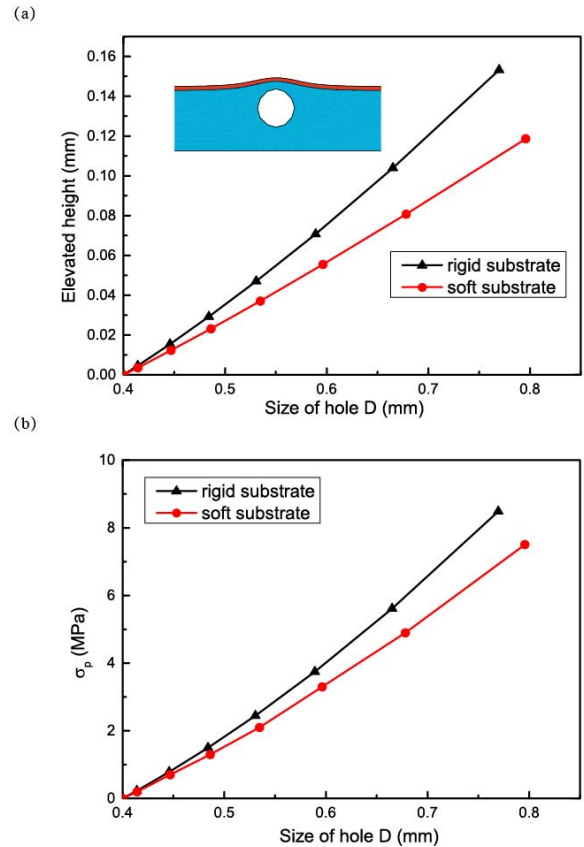

**Figure 2.** (a) Elevated height of pimple bump as a function of the size of follicle for two different substrates; the inset shows the deformation of pimple bump. (b) Mechanical stimulus to mechanoreceptors versus the size of follicle.

The pain is usually perceived during the growth of pimple. The nociceptive pain induced by mechanical stimulus is the focus of our study, though the accumulation of undesirable bacteria in the secreted sebum may cause inflammatory pain [9]. The stretching and compressive strains in both the epidermis and dermis can be used as the stimulus input for the holistic model. We adopt the maximal principle stress at the middle of epidermis (point $A_0$ shown in Figure 1(b)) to characterize the mechanical stimulus, since mechanoreceptors averagely locate at the depth of 50 μm [19,25]. Figure 3(a) shows the predicted membrane voltages of mechanoreceptors under mechanical stimuli of 1.0 MPa and 8.0 MPa, and the generated frequencies are respectively 54 Hz and 112 Hz. Higher level of stimulus induces higher frequency, and the trend is similar with that predicted in the thermal pain model [16]. Furthermore, the corresponding variation of output potential from T-cell with time is shown in Figure 3(b). The voltage increases with time and then approaches a plateau, which is regarded as the pain level. The signal input with frequency 54 Hz induces the high level of voltage, -7.8 mV, while that with frequency 112 Hz causes the output voltage -43.5 mV of T-cell. Since the firing threshold of T-cell is -55 mV, these two signals are able to evoke the sensation of pain. Combined with Figure 3(a), Figure 3(c) clearly shows the increasing frequency response and pain level with the pimple becoming larger, which echoes with our common sense of increased pain perceived during the development of pimple. Accordingly, it's deduced that higher pain level are produced in the pimple with hard substrate than those with soft substrate. The result is exemplified by the perception of higher pain from the pimple grown on the forehead with less subcutaneous fat.

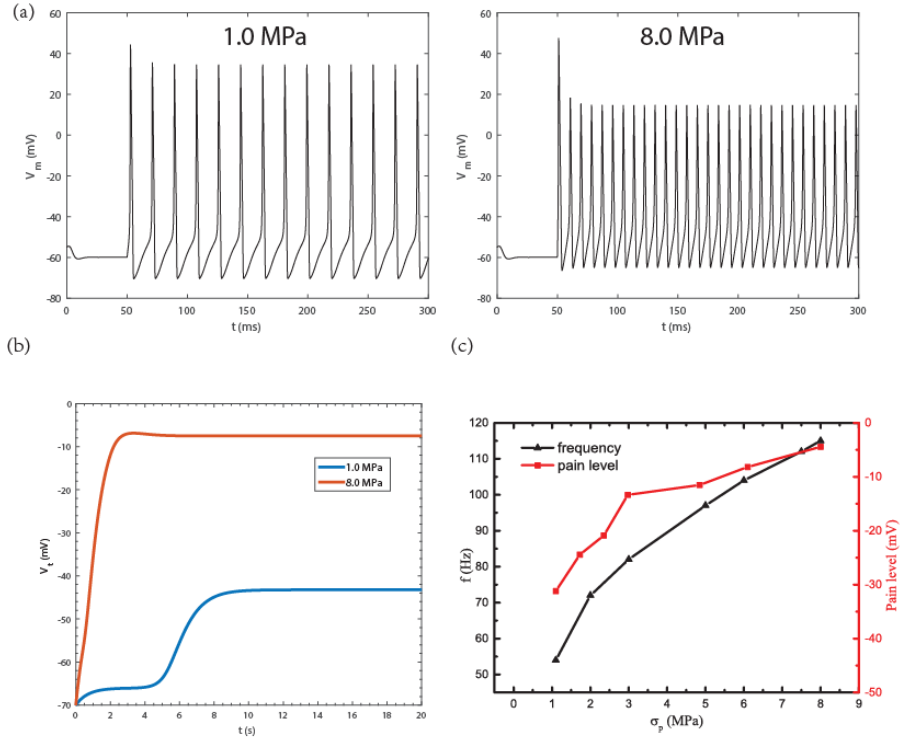

**Figure 3.** (a) The membrane potentials and (b) corresponding output potentials from T-cells under selected stimuli of 1.0 MPa and 8.0 MPa. (c) The frequency of membrane potential and pain level as a function of mechanical stimulus during the growth of pimple.

**3.2 The effect of hole depths and epidermis's property**

Since the depth of sebaceous glands under the skin varies with body sites, the distance, $h$, from the follicle to the surface of skin may play a remarkable role in the growth of pimple [10]. As shown in Figure 4(a), when the follicle is closer to the skin surface, both the bump height and corresponding pain level increase. It's exemplified by the larger elevated height and pain level for the case with hole depth 0.25 mm than those for the one with 0.5 mm. It suggests that higher level of pain is evoked from the pimple grown on the part where sebaceous glands are closer to the skin surface, and various deformations of pimple are exhibited among different body sites.

In addition, some cosmetic products are used to hydrate and moisture skin and may alter the material property of epidermis [26], and they are likely to reduce the pain from pimple to some extent. The stress stimulus and pain level during the growth of pimple for different material properties of the epidermis, 102 GPa for rigid epidermis and 80 GPa for soft epidermis, are plotted in Figure 4(b). As Young's modulus of the epidermis decreases, both mechanical stimulus and pain level decrease. This implies a possible way to relieve the pain level by reducing the stiffness of epidermis.

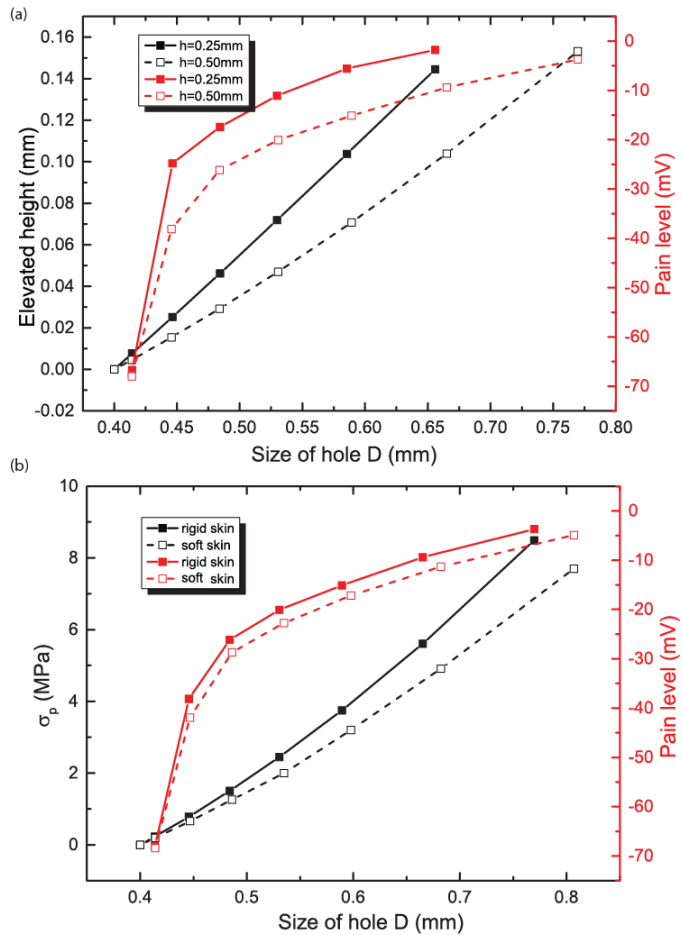

**Figure 4.** The elevated height of pimple bump and pain level as a function of the follicle size for various (a) depths of the follicle and (b) mechanical properties of epidermis.

## 4  Conclusion

Using both the finite element model and holistic method, we explore the deformation induced by a growing pimple and link the stress stimulus to the sensation of nociceptive pain. During the development of pimple, both the height of pimple bump and stimulus to mechanoreceptors increase with the increasing volume of secreted sebum. Different substrates are used to model various environments around pimples, and it is found that the pimple grown on the soft part (e.g., cheeks) leads to smaller stress stimulus and lower pain level compared to that grown on the hard part (e.g., forehead). In addition, the elevated height of bump induced by deeper pimple is lower than that caused by the one closer to the skin surface, and the former evokes smaller pain. Furthermore, when the Young's modulus of epidermis decreases, the stimuli to nociceptors are relatively decreased, which implies that certain kinds of cosmetic products could be used to soothe skins and reduce their mechanical properties.

The actual component and environment that the skin encounters are complex. The present FEM model just qualitatively characterizes a growing pimple, and has not yet considered viscosity and the complicated structure of skin, including the components of hair, capillary network, plexus, etc.[27]. Thus, more efforts in the future should be paid to refine this model by considering the viscoelasticity, anisotropy of skin and the complex components around the pimple. It should also be noted that the plane strain models in this study lead to the similar trends of axisymmetric models, which will be further extended to three-dimensional cases in the future. Additionally, since the bacteria-induced inflammation plays a critical role in the post stage of pimple [9], the holistic method herein considering mechanical stimulus is only reasonable to characterize the sensation of pain at the initial stage of pimple growth. Due to lack of experimental data, we adopt the empirical parameters in the Hodgkin-Huxley model and gate control theory [22,23,28]. The refinement of modeling pain from pimple will be subjected to the future work. Moreover, the results in the current study enable applications on medical and cosmetic method to relieve pain from pimples.

**Acknowledgements**

X.C. acknowledges the support from the National Natural Science Foundation of China (11172231 and 11372241), ARPA-E (DE-AR0000396) and AFOSR (FA9550-12-1-0159); X.L. and H.X. acknowledge the China Scholarship Council for the financial support.

**References**

[1] K. Parsons, *Human thermal environments: the effects of hot, moderate, and cold environments on human health, comfort, and performance* (Crc Press, 2014).
[2] G. Plewig and A. M. Kligman, *Acne: morphogenesis and treatment* (Springer Science & Business Media, 2012).
[3] A. Pearl, B. Arroll, J. Lello, and N. Birchall, The New Zealand medical journal **111**, 269 (1998).
[4] I. Aslam, A. Fleischer, and S. Feldman, Expert opinion on emerging drugs **20**, 91

(2015).

[5] S. Feldman, R. E. Careccia, K. L. Barham, and J. Hancox, American Family Physician **69**, 2123 (2004).

[6] K. Degitz, M. Placzek, C. Borelli, and G. Plewig, Journal der Deutschen Dermatologischen Gesellschaft = Journal of the German Society of Dermatology : JDDG **5**, 316 (2007).

[7] A. Morrison, S. O'Loughlin, and F. C. Powell, International Journal of Dermatology **40**, 104 (2001).

[8] H. C. Williams, R. P. Dellavalle, and S. Garner, The Lancet **379**, 361 (2012).

[9] L. L. Levy and J. A. Zeichner, American journal of clinical dermatology **13**, 331 (2012).

[10] D. Y. Paithankar *et al.*, The Journal of Investigative Dermatology **135**, 1727 (2015).

[11] L. Anderson, *Looking Good: The Australian guide to skin care, cosmetic medicine and cosmetic surgery* (Australian Medical Publishing, 2006).

[12] J. Yin, Z. Cao, C. Li, I. Sheinman, and X. Chen, Proc Natl Acad Sci U S A **105**, 19132 (2008).

[13] J. Yin, G. J. Gerling, and X. Chen, Acta Biomaterialia **6**, 1487 (2010).

[14] L. Zhu and X. Chen, Acta Biomater **9**, 7968 (2013).

[15] J. Van Hees and J. Gybels, Journal of Neurology, Neurosurgery & Psychiatry **44**, 600 (1981).

[16] F. Xu, T. Wen, T. J. Lu, and K. A. Seffen, Journal of biomechanical engineering **130**, 041013 (2008).

[17] F. Xu, T. J. Lu, and K. A. Seffen, Journal of Thermal Biology **33**, 223 (2008).

[18] F. Xu, M. Lin, and T. J. Lu, Computers in biology and medicine **40**, 478 (2010).

[19] F. Xu, T. Wen, K. Seffen, and T. Lu, Applied Mathematics and Computation **205**, 37 (2008).

[20] G. Wei, B. Bhushan, and P. M. Torgerson, Ultramicroscopy **105**, 248 (2005).

[21] N. James and A. Richard,  (Oxford University Press, Oxford, 1996).

[22] J. A. Connor, D. Walter, and R. McKowN, Biophysical Journal **18**, 81 (1977).

[23] A. L. Hodgkin and A. F. Huxley, The Journal of physiology **117**, 500 (1952).

[24] R. MELZACK and P. D. WALL, Survey of Anesthesiology **11**, 89 (1967).

[25] L. Kruger, E. Perl, and M. Sedivec, Journal of Comparative Neurology **198**, 137 (1981).

[26] H. Dobrev, Skin Research and Technology **6**, 239 (2000).

[27] F. H. Silver, L. M. Siperko, and G. P. Seehra, Skin Research and Technology **9**, 3 (2003).

[28] N. Britton, M. Chaplain, and S. M. Skevington, Mathematical Medicine and Biology **13**, 193 (1996).